\documentclass[apjl]{emulateapj}
\usepackage{graphicx}

\shorttitle{A short-period censor of sub-Jupiter mass exoplanets}
\shortauthors{Szab\'o and Kiss}

\begin{document}

\title{A short-period censor of sub-Jupiter mass exoplanets with low density}

\author{
Gy. M. Szab\'o\altaffilmark{1,2},
L. L. Kiss\altaffilmark{1,3}
}
\altaffiltext{1}{Konkoly Observatory of the Hungarian Academy of Sciences, PO. Box 67, H-1525 Budapest, Hungary}
\altaffiltext{2}{Hungarian E\"otv\"os Fellow at the University of Texas at Austin}
\altaffiltext{3}{Sydney Institute for Astronomy, School of Physics A28, University of Sydney, NSW 2006, Australia}





\begin{abstract}

Despite the existence of many short-period hot Jupiters, there is not one hot Neptune with an orbital period less than 2.5 days.
Here we discuss a cluster analysis of the currently known 106 transiting 
exoplanets to investigate {  a possible explanation for} this observation. 
We find two distinct clusters in the mass-density space, one with hot Jupiters 
with a wide range of orbital periods (0.8--114 days)
and a narrow range of planet radii (1.2$\pm0.2$~$R_J$); and 
another one with a mixture of super-Earths, hot Neptunes and hot Jupiters, exhibiting a surprisingly narrow period distribution (3.7$\pm0.8$~days). 
{  These two clusters follow strikingly different distributions in the period-radius parameter plane. The branch of sub-Jupiter mass exoplanets is censored by the orbital period at the large-radius end}: no planets with mass between 0.02--0.8 $M_{J}$ or with radius between 0.25--1.0 $R_{J}$ are known with $P_{orb}<2.5$ days.  
{  This clustering is not predicted by current theories of planet formation and evolution that we also review briefly.}
\end{abstract}


\section{Introduction}

In circumstellar disks with on-going planet formation, protoplanets accrete the gaseous matter of the disk. The initial mass spectrum of planets will depend on the orbital distances via the thermal and tidal effects of the star (Wuchterl et al. 2008). Initial mass functions cover a wide range of masses, exhibit a bimodal or multimodal distribution, peaking near the mass of Neptune and Jupiter. Both peaks move to larger masses with increasing orbital periods. A key feature of the population in the 1--16 day orbital period range is that hot Neptunes always significantly outnumber hot Jupiters (Broeg, 2006) which has been confirmed by observations recently in the 3--100 day period range (Howard et al. 2010). 
The observational confirmation of the mass spectrum is a direct tracer of the physical processes of planet formation in the nebular phase and the further evolution.

As of writing, 106 transiting exoplanets have been published for which precise masses, radii and orbital periods are known (Schneider, 2010). This number of transiting planets is sufficient to perform statistical tests. The most important planet parameters are mass and radius, being the prime tracer of the interior structure (e.g. Guillot 2005, Fortney et al. 2007, Chabrier et al. 2009). Because there are obvious selection effects in the currently known exoplanet sample, tests should either concentrate on unbiased parameters, such as orbital periods, or on the combination of biased parameters, such as masses versus radii.

In this paper we demonstrate that exoplanets in the mass--density space fall into two clusters which follow very different period--radius distributions. Comparison of these samples shows that the observed lack of hot Neptunes and hot sub-Jupiters are censored at $<2.5$ day period range, {  while hot Jupiters do not suffer a similar censor.} We address several candidate scenarios that may explain the observed period--mass distributions.

\begin{figure*}
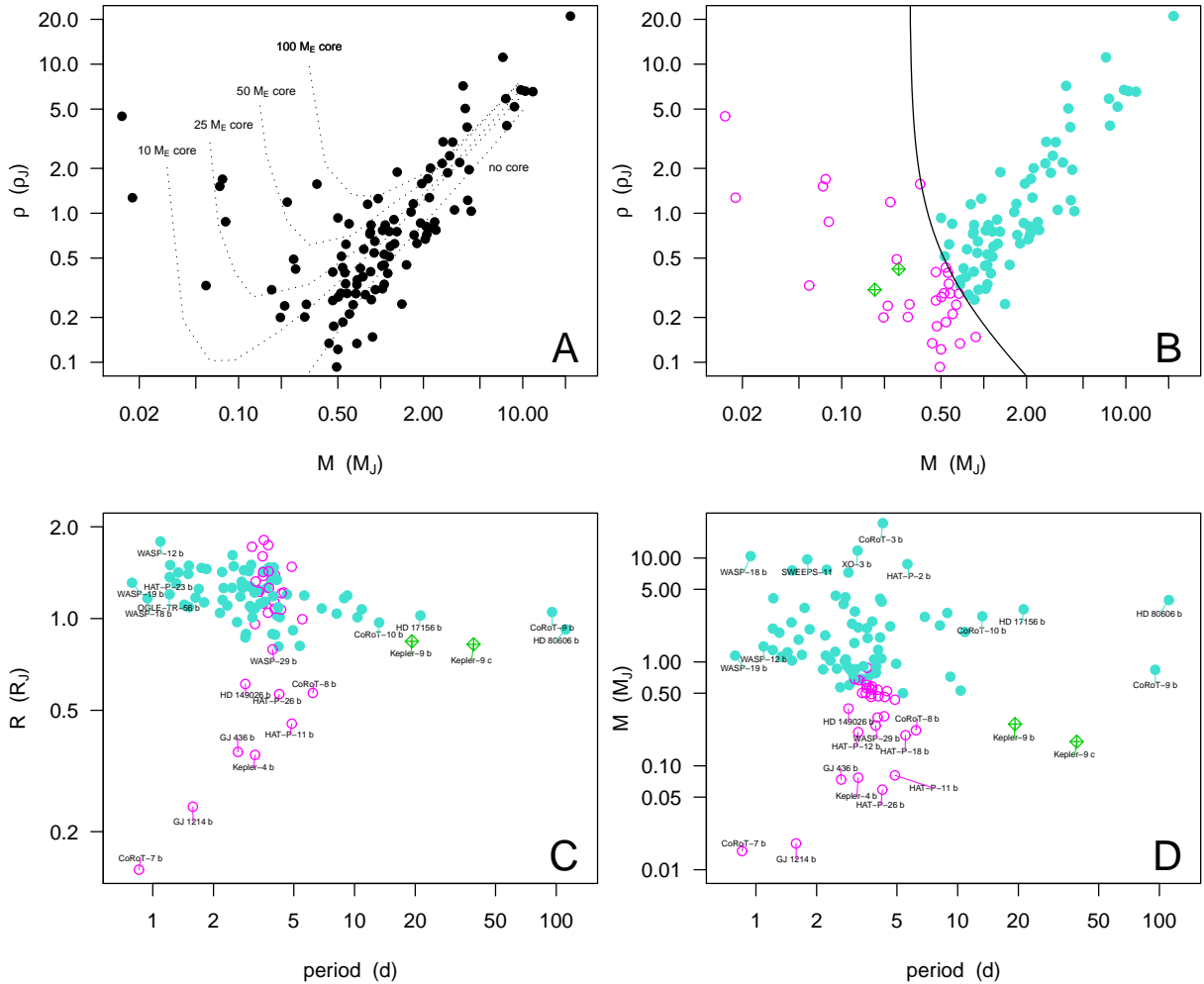

\includegraphics[width=8cm]{golya.epsi}
\includegraphics[width=8cm]{beka1.epsi}
\bigbreak
\includegraphics[width=8cm]{beka2.epsi}
\includegraphics[width=8cm]{beka3.epsi}
\caption{
{  A:} The mass--density diagram of the currently known transiting exoplanets. 
Four planetary models with various core masses and one other without a core are plotted with illustrative purposes. 
{  B:} The proposed clustering in the
mass--density space. Open and filled dots (magenta and blue on-line) distinguish the two clusters. Two planets, Kepler-9 b and c are plotted with diamonds.
{  C:} Clusters form two apparent sequences in the period--radius space. Note the lack of $D_1$ exoplanets with $period<2.2$ days. 
{  D:} The distribution of the cluster members in the period--mass space.}
\end{figure*}

\section{Cluster analysis of the data}

Color-coding the points in one distribution and then plotting another diagram with the specified colors is a powerful technique to find clusters heuristically (see e.g. a nice example in Ivezi\'c et al. 2002 for Solar System asteroids). Existence and structure of the suspected clusters can further be investigated by statistical tools. For such an analysis we collected data from the Interactive Extrasolar Planet Catalog (Schneider, 2010 and references therein). We included mass, size, average density, period, semi-major axis and inclination, and stellar parameters in our analysis. Drawing a few color-coded plots with a trial-and-error strategy, we have found that color-coding from the mass-density space distinguishes two different sequences in the period-radius space. Based on this finding, we propose a discrimination diagram in the MD space, separating two subclasses of exoplanets, hereafter $D_1$ (objects with lower mass) and $D_2$ (objects with larger mass, right to the borderline).

In the panel {  A} of Fig. 1, we plot all the involved transiters in the mass--density space. With purpose of illustration, the 500 Myr isochrone of exoplanets at 0.045 AU distance to a solar analog is indicated (Fortney et al. 2007). The models contain 50\% silicate and 50\%{} water core with various initial masses indicated by the labels, and total mass indicated in the abscissa. 

The border between the two clusters (panel {  B} of Fig. 1) was determined by maximizing the Mahalanobis distance (Mahalanobis, 1936) of the two distributions in the period-radius space (see details in Appendix). Mahalanobis distance has been successfully used in clustering data with severe correlations (De Maesschalk, 2000). The borderline between the two clusters has the equations of 
$\log(\rho) = 0.12513 \log(M-0.30779)^{-0.83645}$, cluster $D_1$ is to the left (open symbols in panel {  B} pf Fig. 1), while $D_2$ is to the right. 

This classification can in part be assigned to objects discriminated by mass. $D_1$ exoplanets
contain super-Earths (Valencia et al. 2007, Adams et al. 2008), hot Neptunes, and low-mass, low-density hot Jupiters. Hot Jupiters exceeding the mass or the density of Jupiter are assorted in $D_2$. In sake of a suggestive distinction, $D_1$ may be referred as sub-Jupiters though cluster membership slightly depend on density, too. For example, a planet with 0.6 Jupiter mass will be $D_2$ if its density is around that of Jupiter and will be $D_1$ if the density is significantly lower.

\section{Results}

The defined clusters (panel {  B} of Fig. 1) are well separated in the period--radius and period--mass parameter planes (panels {  C} and {  D}). $D_2$ cluster members cover a period range of more than 2 orders of magnitude, beginning from 0.7 days. The radius of $D_2$ members are indicatively 1--2 times that of Jupiter, slightly decreasing with orbital period (panel {  C} of Fig. 2, solid dots). The dependence of radius on period is due to the sensitivity of planet atmospheres to stellar irradiation (Fortney et al. 2007). 

Transiters in $D_1$ cluster exhibit a narrow period distribution: 85\%{} of them lie in the 2.5--5 days range. {  The branch of} massive $D_1$ members overlap with $D_2$ cluster members {  at the large-radius end} in the period--radius space. {  All D1 cluster members have periods $>2.88$ days  in the $R>0.5$~$R_J$ size range, while there are 33 hot Jupiters in the 0.79--2.88 day orbital period range. Because these planets have similar radius and only their density differs, we conclude that the density is a major parameter in relation to the period censor.}

Kepler 9b and c (Holman et al. 2010) are two important outliers with orbital periods of 19.2 and 38.9 days. However, these planets exhibit prominent variation of the period together with transit timing variations as signatures of gravitational interaction of two planets near the 2:1 orbital resonance (Holman et al. 2010). Because of this known instability, designated them with diamond symbols (green on-line).

In the lower two panels of Fig. 1, a period desert of $D_1$ exoplanets is apparent, outlined by planets CoRoT-7b, GJ 1214 b, GJ 436 b, and HD149026 b. This means that no sub-Jupiters have orbital period less than 2.5 days, except the super Earths. In the followings we estimate the probability of a low-period desert in the $D_1$ sample to occur by chance. There are 2 of 31 (6.5\%{}) $D_1$ exoplanets with a period less than 2.5 days. The upper boundary of the period range of $D_1$ exoplanets is 7 days. We compare the period distribution of $D_1$ exoplanets to hot Jupiters which have period less than 2.5 days. There are 64 transiting hot Jupiters within this period range, and 24 of them (38\%{}) have a period less than 2.5 days. 

The two-sided Fisher's exact test (Fisher 1925, Douglas 1976) is a statistical test used to determine if there are nonrandom associations between two categorical variables. 
Testing the above contingencies, the asymmetric distribution is confirmed at the 99\%{} confidence level, which is an evidence for the presence of a low-period desert in the data. {  The period censor does not affect large
density hot Jupiters which is evidently seen at the large-radius end of the $D_1$ distribution, where it overlaps with $D_2$s. The censor may be density selective, and this would be the primary cause of the ``hole'' in the period--radius distribution.}

\begin{figure}
\includegraphics[width=\columnwidth]{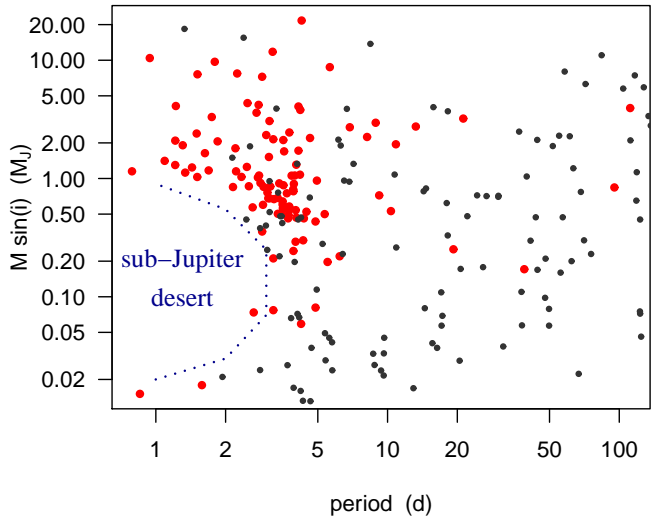}
\caption{The period--minimum mass diagram of the currently known exoplanets. Large (red) dots: transiters; small (grey) dots: non-transiting planets. The axis ranges fit the panel {  D} of Fig. 1.}
\end{figure}

The period--mass diagram can be completed with non-transiting exoplanets. In this case the minimum mass, $M\sin i$ is known, and mass information is ambiguous to some extent. This is not a serious problem because the sub-Jupiter desert has an extension of 1.5 orders of magnitude in mass, and the hole in the period--mass distribution will remain recognizable. In Fig. 2 we plot minimum mass versus period, using different symbols for transiters and non-transiting planets. Because of the strategy of radial velocity surveys, the period distribution of non-transiting planets is Nyquist-limited at around $P\approx 2$ day frequency, and there they exhibit not much overlap with the sub-Jupiter desert. However, there are 9 non-transiting planets with orbital periods shorter than 3 days, falling around the long-period edge of the desert, but none coinciding with the desert itself. {  The lack of sub-Jupiters with semi-major axes less than 0.03 AU, or at least an anticorrelation between mass and semi-major axis was suggested by previous authors (e.g. Winn et al. 2010). It has to be noted that the ``hole'' is more apparent in the period-radius distribution that we plot in this paper. Moreover, we concluded that there are important sub-structures in the distribution: nearly zero or slightly positive correlation exists between the period and the radius of D1 exoplanets, and a very slight anticorrelation exists for hot Jupiters.}

\section{Discussion}

In contradiction with the current models of planet formation nearby a star (e.g. Broeg et al. 2009), we found that sub-Jupiter exoplanets (more exactly: the $D_1$ cluster members in the mass-density parameter space) exhibit a desert for $P_{orb}<2.5$ day periods. The observed distribution is inconsistent with models assuming that most planets are born near or beyond the snow-line and then migrate inwards (Ida \&{} Lin, 2008, Mordasini et al. 2009). These models lead to a super-Earth desert predicting a paucity of planets in the mass range of 1--30 $M_{Earth}$ and orbiting inside 1 AU. Instead, our analysis showed that there are many sub-Jupiters between 3--5 day orbital period, which confirms the conclusion of Howard et al. (2010), telling that the current population synthesis models are inadequate to explain the distribution of low-mass planets.

Discussing possible modifications of planet formation models lies out the scope of this work, here we suggest some possible hypotheses to account for.
\medbreak

{  1)} Hot Neptunes may evaporate rapidly in the close vicinity of the star. Highly exposed planets with less potential energy evaporate more rapidly (Lecavelier des Etangs, 2007) which can be a candidate explanation for the period censor. An argument for this procedure is the dependence of $D_1$--$D_2$ clustering on the density: planets with loose atmospheres are classified as $D_1$ in the mass range of 0.5--1 $M_{J}$. Indeed, these are the planets which exhibit the period censor.

However, the presence of short-period super-Earths raises questions in regards of the evaporation framework. The internal energy of their atmosphere, if there is any, is evidently low and they should evaporate more rapidly than Neptunes. In the contrary, the hot super-Earth GJ 1214 is considered to have a thick atmosphere (Rogers and Seager, 2010). {  A possible explanation could be that the incident UV flux is low around the host M dwarf star and the atmosphere can survive.}

{  2)} As sufficient explanation for the low period desert, one could assume that hot Neptunes spiral in or migrate outwards in a short time-scale, while hot Jupiters do not. An argument against the selective in-spiraling of hot Neptunes is found by Armitage (2007), concluding that the mass function is not affected significantly by migration. Since many hot Jupiters with orbital periods of 0.7--2.5 days survived type II/III migration, it may be difficult to explain why hot Neptunes did not. Tidal disruption cannot explain the censor either, because hot super-Earths with 1 day orbital periods are still stable against tidal disruption (Schlaufman et al. 2010). Hot Neptunes can migrate outwards (Martin et al. 2007), which process could also have evacuated the sub-Jupiter desert. However, outward migration requires a massive inner planet (Martin et al. 2007) which lacks in the case of the known hot Neptunes.

{  3)} Based on their completely different distributions in the period--mass diagrams, one could assume that $D_1$ and $D_2$ planets differ in the amount and strength of planet-planet perturbation. However, this idea is generally challenged by the distribution of eccentricities and stellar obliquity. Both $D_1$ and $D_2$ cluster exoplanets share similar eccentricity properties, probably reflecting similar perturbation history. To check this, we applied a Kolmogorov-Smirnov test to the eccentricities of $D_1$ and $D_2$ members, resulting in a $p=0.55$ value, suggesting that the eccentricities are very similarly distributed.
Stellar obliquities follow a similar distribution in $D_1$ and $D_2$, too. There are nine systems with large stellar obliquity known among $D_2$ exoplanets and two are known in the $D_1$ cluster. Confirmed by a Fischer's Exact Test, there is no significant difference in the occurrence of large obliquity.

{  4)} The observed long-period edge of the sub-Jupiter desert fits the type I migration model predictions of Masset et al. (2006) with disk torques accounted. Near the disk cavity, a density radial jump forms in the protoplanetary nebula. In the theory of Masset et al. (2006), low-mass objects reaching the disk cavity will be trapped and halt migrating. The radius of the jump highly depends on the structure and the density of the disk, and it is about 0.03 AU for a disk having the same surface density profile as the minimum mass solar nebula. This scenario correctly predicts the 2--2.5~day period censor of sub-Jupiters migrating in a tenuous disk environment. However, this scenario cannot explain the bottom edge of the sub-Jupiter desert, i.e. the presence of short-period hot super-Earths -- which should also have been trapped at the disk cavity. To resolve the contradiction with observations, an appropriate modification is neccessary to explain low-mass ($M<$0.02~$M_{Jup}$) exoplanets on $P<2.5$ day orbits.

\section*{Acknowledgments}

This project has been supported by the Hungarian OTKA Grants K76816, 
K83790 and MB08C 81013, the ``Lend\"ulet'' Program of the Hungarian
Academy of Sciences. GyMSz was supported
by the ``E\"otv\"os'' Fellowship of the Hungarian Republic.

\begin{appendix}

\section{Clustering with maximal Mahalanobis distance}

The Mahalanobis distance of a generic vector $x$ and the $\mu_1$ barycenter of a given distribution, $D_1$, is calculated with accounting for the coordinate correlations of $D_1$:
\begin{equation}
d_M({x},D_1) = 
  \sqrt{(x-\mu_1)^T \ \ S_{D1}^{-1} \ \ (x-\mu_1)}, 
\end{equation}
where $S_{D1}$ is the covariance matrix of $D_1$, and acts as the metric tensor in this definition. We define the Mahalanobis distance of $D_1$ and $D_2$ clusters as the sum of Mahalanobis distances of all points in $D_1$ from $D_2$, plus that of all points is $D_2$ from $D_1$:
\begin{equation}
d_M(D_1,D_2) = \sum_{\forall x_1 \in D_1} d_M(x_1,D_2) + \sum_{\forall x_2 \in D_2} d_M(x_2,D_1).
\end{equation}
The discrimination curve was defined in the MD space, allowing for a curvature as
\begin{equation}
\log(\rho) = c_1 \log(M-c_2)^{c_3},
\end{equation}
where $c_{1,2,3}$ are free parameters to be fitted, $\rho$ is the average density and $M$ is the mass of exoplanets. 
In fact, maximizing $d_M(D_1,D_2)$ does not lead to the desired result because it converges to one large cluster and another single outlier. The quantity to be optimized for is the increment of the Mahalanobis distance due to the clustering, i.e.
\begin{equation}
d_M(D_1,D_2)\ -\ \big\langle\ d_M(F_1,F_2)\ \big\rangle,
\end{equation}
where $F_1$ and $F_2$ are disjunct clusters randomly selected from the whole sample by elements, and they have the same amount of elements as $D_1$ and $D_2$. Since there are numerical fluctuations in $d_M(F_1,F_2)$ due to the stochastical selection, Mahalanobis distances of many random clusterings must be calculated and averaged (which is represented by the $\langle\rangle$ symbols, here standing for the expectation
value). Via altering $c_{1,2,3}$, the $D_1$--$D_2$ clustering varies and in such way the discrimination in the MD space can be optimized for the maximal Mahanalobis distance in the PR space.

In our calculus, initial parameters were $c_{1,2,3}=0.13,0.3,0.85$, clustering 33 objects to $D_1$. Mahalanobis distance was minimized with a random walk algorithm, altering the initial parameters by a factor randomly distributed normally with 1 expectation value and 1.5\%{} FWHM. When the clustering converged, in total 27 exoplanets remained in the $D_1$ cluster.

\end{appendix}

\label{lastpage}

\end{document}